\documentclass[traditabstract]{aa} % for a referee version
\usepackage{multirow}
\usepackage{float,lscape}
\usepackage{rotating}
\usepackage{varwidth}
\usepackage{textpos}
\usepackage{rotating} % Rotating table
\usepackage{graphicx, epstopdf}
\usepackage{natbib}
\usepackage{amsmath}
\usepackage{mwe,tikz}
\usepackage{amssymb}
\usepackage[percent]{overpic}

\usepackage[para,online,flushleft]{threeparttable}
\usepackage{txfonts}
\usepackage{hyperref}
\def\inte{{\em INTEGRAL}}

\def\chan{{\em Chandra}}

\def\rxte{{\em RXTE}}
\def\swift{{\em Swift}}

\def \inte {{\em INTEGRAL}}

\def \nustar {{\em NuSTAR}}

\def \nicer {{\em NICER}}

\def \ferg {erg cm$^{-2}$ s$^{-1}$}
\def \hcm {\hbox {\ifmmode $ atom cm$^{-2}\else atom cm$^{-2}$\fi}}

\def \igr {IGR\,J17503-2636}

\begin{document}

   \title{IGR\,J17503-2636: a candidate supergiant fast X-ray transient}

   \author{C. Ferrigno
    \inst{1}
    \and E. Bozzo
    \inst{1}
    \and A. Sanna
    \inst{2}
    \and G. K. Jaisawal
    \inst{3}
    \and J. M. Girard
    \inst{1}
    \and T. Di Salvo
    \inst{4}
    \and
    L. Burderi
    \inst{2}
    }
    
   \institute{Department of Astronomy, University of Geneva, Chemin d'\'Ecogia 16, CH-1290 Versoix, Switzerland; \email{carlo.ferrigno@unige.ch}
    \and 
    Universit\'a degli Studi di Cagliari, Dipartimento di Fisica, SP Monserrato-Sestu, KM 0.7, 09042 Monserrato, Italy
    \and 
    National Space Institute, Technical University of Denmark, Elektrovej 327-328, DK-2800 Lyngby, Denmark
    \and
    Universit\`a degli Studi di Palermo, Dipartimento di Fisica e Chimica, via Archirafi 36, 90123 Palermo, Italy
}

   \date{}

  \abstract{IGR\,J17503-2636 is a hard X-ray transient discovered by \inte\ on 2018 August 11. This was the first ever reported X-ray emission from this source. Following the discovery, follow-up observations were carried out with \swift,\ \chan,\ \nicer,\ and \nustar.\ We report in this paper the analysis and results obtained from all these X-ray data. Based on the fast variability in the X-ray domain,  the spectral energy distribution in the 0.5--80\,keV energy range, and the reported association with a highly reddened OB supergiant at $\sim$10~kpc, we conclude that \igr\ is most likely a relatively faint new member of the supergiant fast X-ray transients. Spectral analysis of 
  the \nustar\ data revealed a broad feature in addition to the typical power-law with exponential roll-over at high energy. This can be modeled either in emission or as a 
  cyclotron scattering feature in absorption. If confirmed by future observations, this feature would indicate that \igr\ hosts a strongly magnetized neutron star with $B$$\sim$2$\times$10$^{12}$~G.}   
  

   \maketitle

\section{Introduction}
\label{sec:intro}

\igr\ is an X-ray transient discovered by the JEM-X instruments \citep{lund03} on-board \inte\ \citep{winkler03AA} during the  
Galactic Bulge observations performed on 2018 August 11 \citep[satellite revolution 1986;][]{chenevez18}. 
At discovery, the source flux in the JEM-X data covering from 07:44 to 12:30 UT was estimated at 
3.9$\times$10$^{-10}$~\ferg (3--19\,keV). The source was not initially detected in the quick-look analysis of 
the higher energy \inte\ instrument IBIS/ISGRI \citep{ubertini03,lebrun03}.   

A 1\,ks-long follow-up observation with \swift/XRT \citep{burrows05} was carried out on 2018 August 13 at 19:41 UT 
in order to provide the first characterization of the source emission in the soft X-rays and 
improve the X-ray position down to arcsec accuracy. The preliminary analysis of XRT data  
revealed a variable absorption in the range (6.2--13.7)$\times$10$^{22}$\,cm$^{-2}$ 
(assuming a Galactic absorption column density in the direction of the source of 1.3$\times$10$^{22}$\,cm$^{-2}$) and 
provided a measurement of the variable 0.3--10\,keV flux during the observation in the range (2.0--8.3)$\times$10$^{-11}$\,\ferg 
\citep{chenevez18}. The most accurate localisation of the source was obtained from a 1~ks-long \chan\ observation at RA(J$2000)$ = $17^{\rm h} 50^{\rm m} 17\fs99$, Dec(J$2000)=-26^{\circ} 36^{\prime} 16\farcs7$ 
with an associated uncertainty of 0\farcs7 at 90\,\% confidence level \citep[the observation was carried out on 2018 August 23 at 23:31 UT;][]{chakrabarty18, chakrabarty18b}. Only eleven events from the source were recorded by \chan,\ and no spectral or timing analysis could be performed (the estimated X-ray flux in the 0.5-10~keV energy band was 2.9$\times$10$^{-12}$~\ferg). The accurate \chan\ position allowed \citet{masetti18} to identify the IR counterpart of \igr\ as an heavily reddened OB (super)giant star located beyond the Galactic Center at about 10~kpc, thus classifying the \inte\ transient as an high mass X-ray binary \citep[HMXB;][]{walter15}. The fast flaring behavior and the rapid decay in the X-ray flux suggests an association of \igr\ to the HMXB sub-class of the supergiant fast X-ray transients \citep[SFXTs;][]{Sguera2005,Negueruela2006,Sguera2006,nunez17}, as we discuss in  Sect.~\ref{sec:discussion}. 
 
In this paper, we report on all available X-ray data that were collected during the first reported X-ray emission episode from \igr\ with the 
instruments on-board \inte,\ \nustar,\ \swift,\ and \nicer,\ together with our interpretations.
 
\section{X-ray data}
For all instruments, we performed spectral analysis with Xspec 12.10.0c \citep{Arnaud96}.  All uncertainties in the paper are given at 90\% confidence level, unless stated otherwise. A summary of all observations with common spectral results is reported in Table~\ref{tab:summary}; we refer to the following paragraphs for a detailed description of the spectral analysis for each X-ray facility.

\begin{table*}
	\scriptsize
	\centering
	\caption{Log of X-ray observations with the best-fit spectral parameters.}
	\label{tab:summary}

	\begin{tabular}{l cccc r@{}l r@{}l c r@{}l c}
		\hline
		\hline
		TELESCOPE & START & STOP & EXP. & OBSID & \multicolumn{2}{c}{$n_\mathrm{H}$} & \multicolumn{2}{c}{$\Gamma$} & {PL Flux (2--10\,keV)\tablefootmark{a}} & Cstat-$\chi^2_\mathrm{red}$ & /d.o.f.\tablefootmark{b} & Flux (2--10\,keV)\tablefootmark{c} \\
		 & UTC & UTC & ks &  & \multicolumn{2}{c}{$10^{22}\,\mathrm{cm}^{-2}$} & \multicolumn{2}{c}{} & {$10^{-11}\mathrm{erg\,cm^{-2}s^{-1}}$} & \multicolumn{2}{c}{} & $10^{-11}\mathrm{erg\,cm^{-2}s^{-1}}$ \\
		 \hline
		\inte & 2018-08-10 15:52 & 2018-08-11 19:20 & 54.5\tablefootmark{d} & N/A & \multicolumn{2}{c}{$<$73} & 2.8&$^{+0.7}_{-0.4}$ & 63$^{+103}_{-35}$ & 0.6 & /10 & 70 \\
		\swift/XRT & 2018-08-13 19:43 & 2018-08-13 20:13 & 0.99 & 00010807001 & 13.0 &$\pm$5.0 & 0.5 &$\pm$0.6 & 14$^{+3}_{-2}$ & 46.0 & /53 & 8.9 \\
		\nicer & 2018-08-14 21:23 & 2018-08-14 22:43 & 0.76 & 1200320101 & 26.0 & $\pm$5.5 & 2.4 & $\pm$0.6 & 8.5$^{+4.8}_{-2.4}$ & 175.0 & /131 & 2.1 \\
		\nicer & 2018-08-16 02:14 & 2018-08-16 13:19 & 2.89 & 1200320103 & 5.0 & $\pm$1.3 & 0.4 & $\pm$0.3 & 1.6$^{+0.1}_{-0.1}$ & 117.0 & /141 & 1.3 \\
		\nustar\tablefootmark{e} & 2018-08-22 11:23 & 2018-08-23 10:29 & 39.1 & 90401333002 & 21.0 & $\pm$2.5 & 0.49 & $\pm$0.12 & 4.1$^{+0.4}_{-0.2}$ & 1.0258 & /324 & 2.3 \\
		\swift/XRT & 2018-08-23 04:39 & 2018-08-23 06:38 & 1.85 & 00088805001 & 56 & $^{+27}_{-22}$ & 2 & $\pm$2 & 9.6$^{+31.0}_{-5.3}$ & 42.0 & /48 & 1.5 \\
		\hline
	\end{tabular}
\tablefoot{
	\tablefoottext{a}{Intrinsic flux of the pegpowerlaw component in the 2--10\,keV energy range.}
	\tablefoottext{b}{For \inte\ and \nustar, we provide the reduced $\chi^2$, for \nicer\ and \swift/XRT, the C-stat.}
	\tablefoottext{c}{Absorbed model flux in the 2--10\,keV energy range.}
	\tablefoottext{d}{This is the IBIS/ISGRI effective exposure corrected for dead time, while each JEM-X unit has an exposure of 6\,ks.}
	\tablefoottext{e}{The reported parameters are for the absorbed power-law component only and model \texttt{Gabs} in Table~\ref{tab:spec} (we refer to  Sect.~\ref{sec:nustar} and Table~\ref{tab:spec} for the complete \nustar\ analysis).}	
}
\end{table*}

\subsection{\inte\ data}
\label{sec:integral}

\inte\ observations are divided into science windows (ScWs), 
i.e., pointings with typical duration of $\sim$2--3\,ks. 
To minimize calibration uncertainties and maximize the exposure, we have selected all publicly available 
pointings with a limited off axis angle from the source: 10~deg from IBIS/ISGRI and
4~deg for JEM-X\footnote{http://www.isdc.unige.ch/integral/analysis}.
All data were processed with the version 11.0 of the Off-line Scientific Analysis software   
(OSA) distributed by the ISDC \citep{courvoisier03}.

\begin{figure}
   \resizebox{\hsize}{!}{\includegraphics{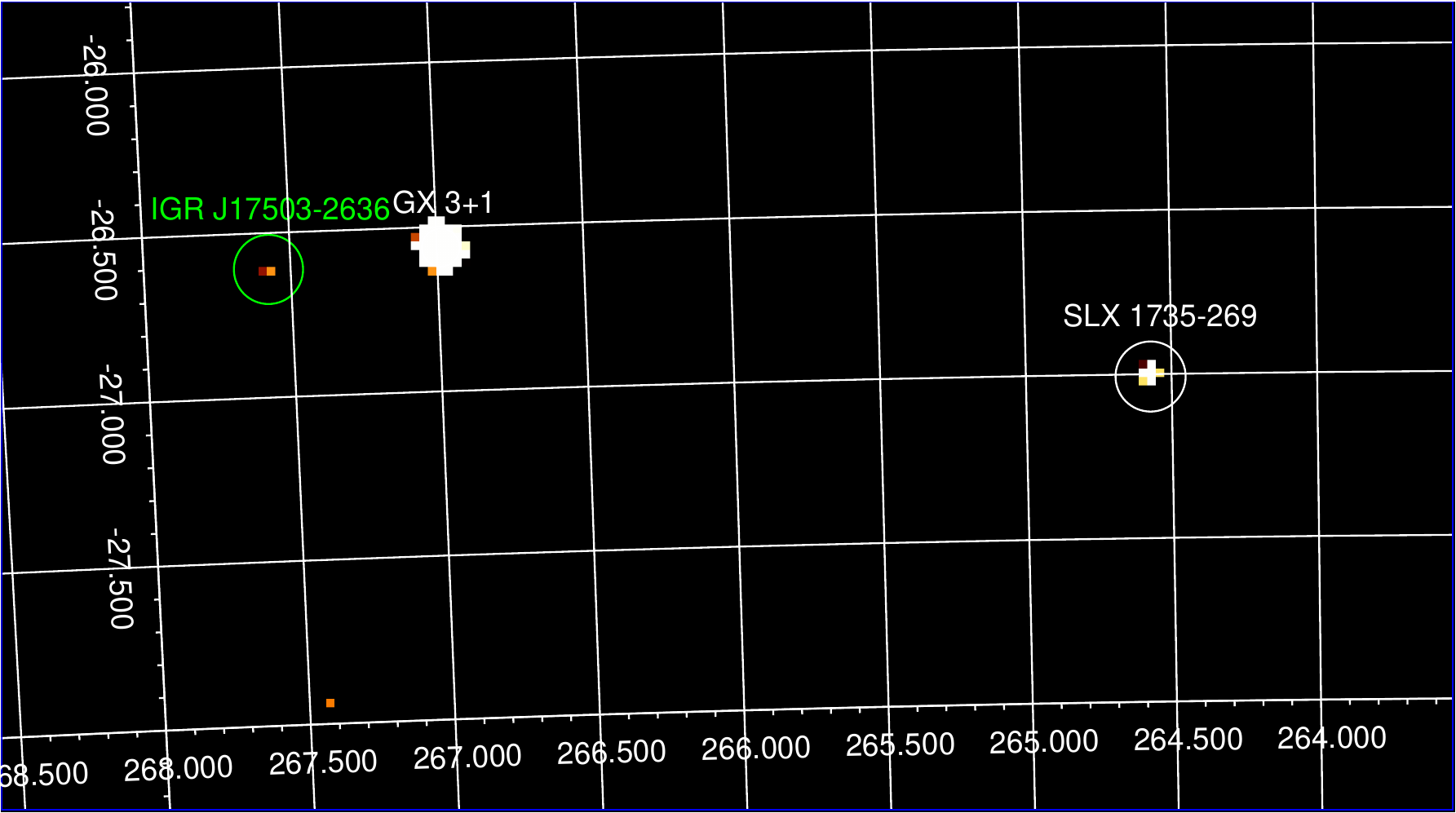}}
    \resizebox{\hsize}{!}{\includegraphics{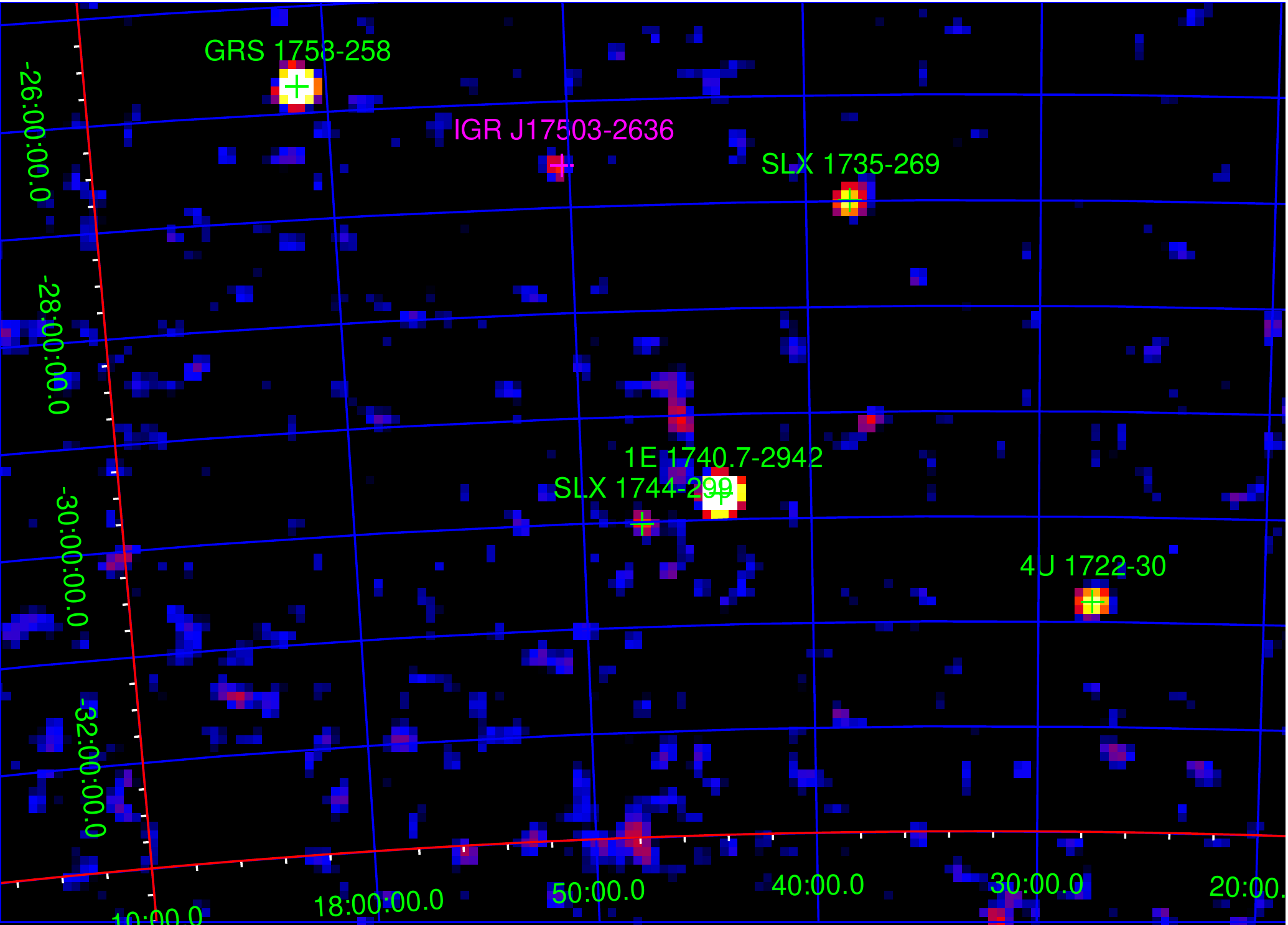}}
  \caption{Zoom of the JEM-X (upper panel) and IBIS/ISGRI (bottom panel) mosaic of the region around \igr\ obtained by combining all 
  publicly available data from from 2018-08-10 at 15:50 to 2018-08-11 at 18:52. The source is detected at a significance of $7\sigma$ in the IBIS/ISGRI 
  mosaic and $6\sigma$ in the JEM-X mosaic.}    
  \label{fig:mosa}
\end{figure}

\begin{figure}
  \resizebox{\hsize}{!}{\includegraphics{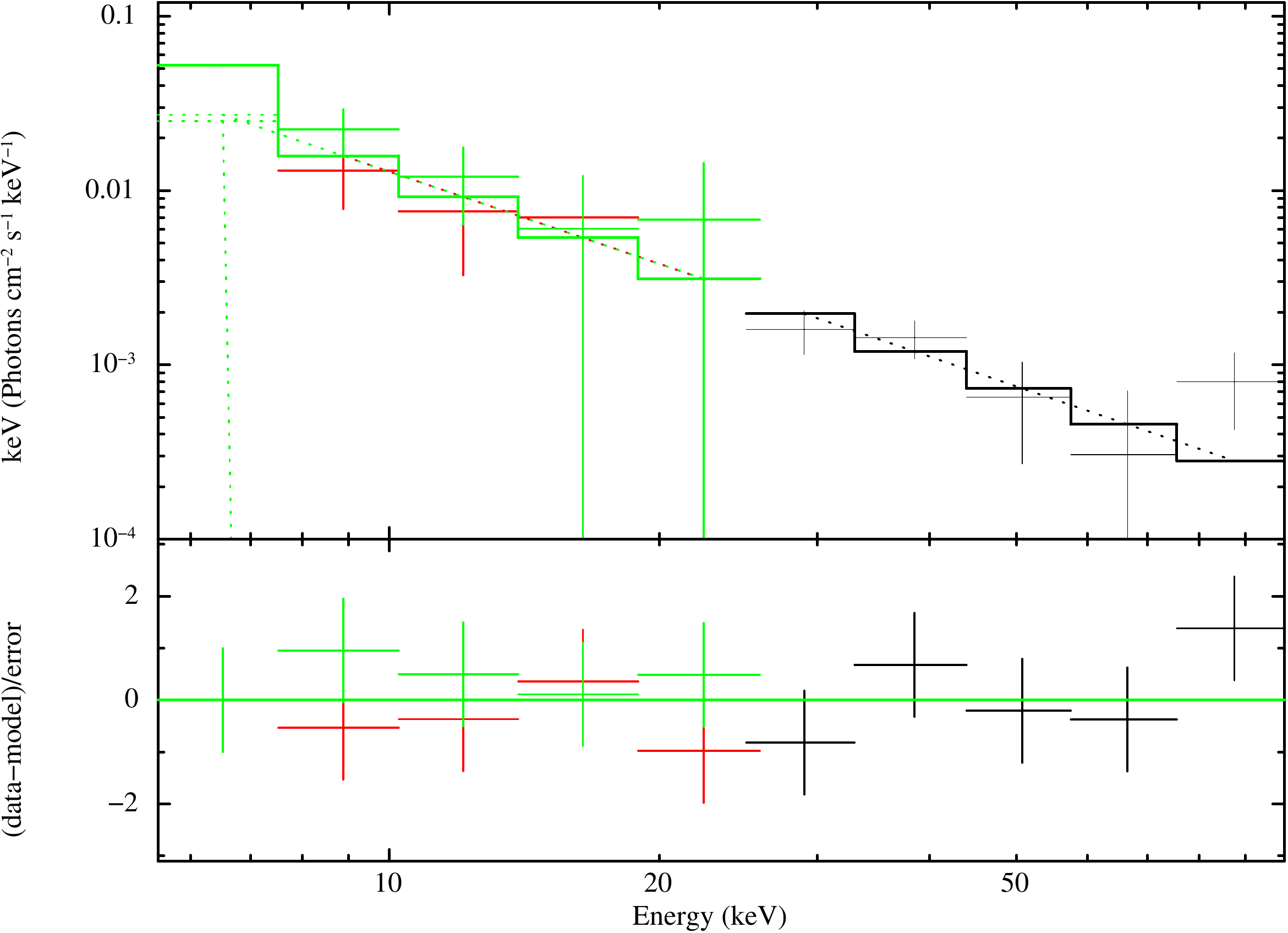}}
  \caption{Unfolded JEM-X1 (red), JEM-X2 (green), and IBIS/ISGRI (black) spectra obtained by integrating over all 
  available data between 2018-08-10 at 15:50 and 2018-08-11 at 18:52. The best fit  
  is obtained with an absorbed power-law model (see text for details). Residuals from the fit are 
  shown in the bottom  panel.}     
  \label{fig:int_spe}
\end{figure}

We extracted the IBIS/ISGRI and JEM-X mosaics by stacking all available data 
from 2018-08-10 at 15:50 to 2018-08-11 at 18:52 together.
\igr\ was detected in the IBIS/ISGRI 25--80\,keV mosaic at a significance of 
$7\sigma$ (single trial, effective exposure time 54.5~ks) and in the JEM-X 3--25\,keV mosaic at a significance of $6\sigma$ (single trial, effective exposure time 21.7\,ks). 
We show a zoom of the IBIS/ISGRI and JEM-X mosaics around the position of \igr\ in Fig.~\ref{fig:mosa}.
We extracted the IBIS/ISGRI light curve in the 25--80\,keV energy range and with one science window time granularity, 
but we could not detect any significant variability (at 3$\sigma$ confidence level). The JEM-X coverage 
(considering a maximum off-axis angle of four degrees) is limited to a few SCWs, not allowing for any variability 
study. 
Thus, we  extracted a single spectrum integrating over the entire exposure time available for ISGRI,  JEM-X1, and JEM-X2 data.
The JEM-X (IBIS/ISGRI) spectra were computed in eight (five) logarithmic equally spaced bins between 3 and 35\,keV (25 and 100\,keV).
Due to instrumental systematic uncertainties and inconsistencies between the two JEM-X units, we have limited the use of data in the range $\sim$5--25\,keV for JEM-X2 and 
$\sim$7--25\,keV for JEM-X1.  
These spectra (Fig.~\ref{fig:int_spe}) could be well fit ($\chi_{\rm red}^2$/d.o.f.=0.6/10) with a simple power-law model (\texttt{TBabs*pegpwrlaw} in {\sc Xspec}) 
with best-fit parameters reported in Table~\ref{tab:summary}.
%We measured a power-law photon index of $2.8^{+0.6}_{-0.3}$ and a 3--80 keV X-ray flux of  $6^{+5}_{-2}\times$10$^{-10}$~\ferg. 
%The absorption column density was loosely constrained in the fit to be $<$2$\times$10$^{23}$\,cm$^{-2}$,
%in agreement with the early \nicer\ observation, see Sect.~\ref{sec:nicer}.  

We have finally checked that during the following INTEGRAL visibility period 
(2018-08-17 14:03-2018-08-19 18:08), the source was not detected. We derived an upper limit on its X-ray flux of $4\times10^{-10}$\ferg\ in the 3--80\,keV energy range at 3$\sigma$ 
confidence level (assuming a power-law photon index of 2.8).

\subsection{\nustar\ data}
\label{sec:nustar}

\igr\ was observed by \nustar\ \citep{nustar} from 2018 August 22 at 11:01 to August 23 at 10:26 (UT; ID~90401333002). 
After having applied to the \nustar\ data all the good time intervals (GTI) accounting for the Earth occultation and 
the South Atlantic Anomaly passages, we obtained an effective exposure time of 39.1~ks for both the  
focal plane modules A and B (FPMA and FPMB). All data were processed via {\sc nupipeline v0.46} and the latest 
calibration files available at the time of writing (v.20181022). The source spectra and light curves were extracted from a 80~arcsec circle centered on the source, 
while the background products were extracted from a region with a similar extension but centered on a region free from the contamination 
of both straylight and source emission. Various extraction regions were also used for the source and background products to verify that 
none of the timing and spectral features could be affected by some specific choices. 

\begin{figure}
 \centering
 \resizebox{\hsize}{!}{\includegraphics{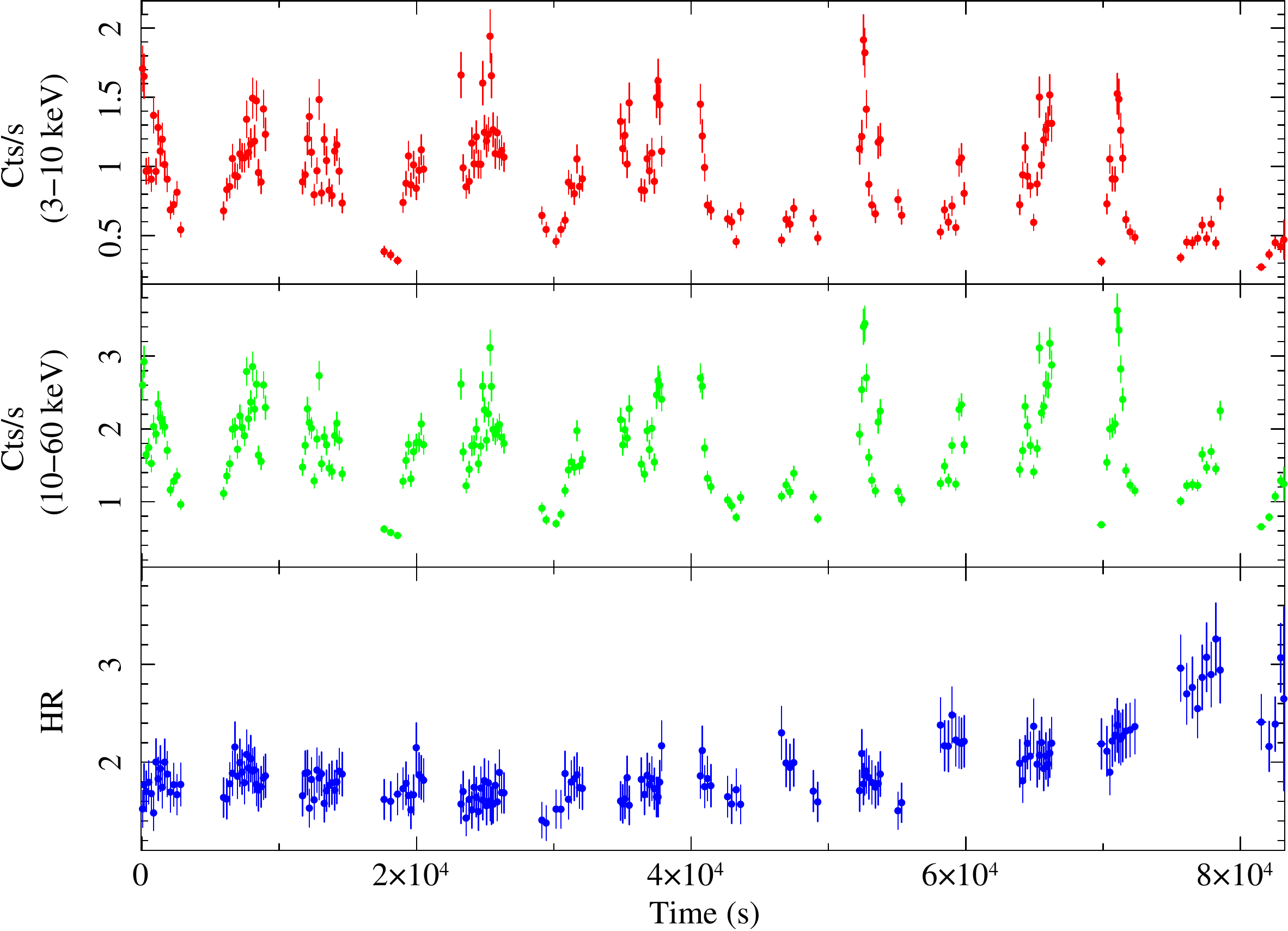}}
  \caption{\nustar\ FPMA lightcurves of \igr\ in the 3--10\,keV and 10--60\,keV energy bands (upper panels) and 
  	the hard to soft ratio (HR, bottom panel). The lightcurves have been rebinned in order to achieve S/N$\gtrsim$10 in each time bin of the soft energy band.}
  \label{fig:nustar_lc}
\end{figure}

The FPMA and FPMB light curves of the source display a remarkable variability. 
We show in Fig.~\ref{fig:nustar_lc} the FPMA lightcurves of the source in two energy bands  
(3--10\,keV and 10--60\,keV) and the corresponding hardness ratio (HR) calculated with an adaptive rebinning of the lightcurves in order to achieve a signal-to-noise ratio 
(S/N) of at least 10 in each time bin \citep[based on the soft lightcurve; see][for more details]{bozzo13}. We verified that compatible results 
could be obtained from the FPMB lightcurves.    

\begin{figure}
	\centering
	\resizebox{\hsize}{!}{\includegraphics{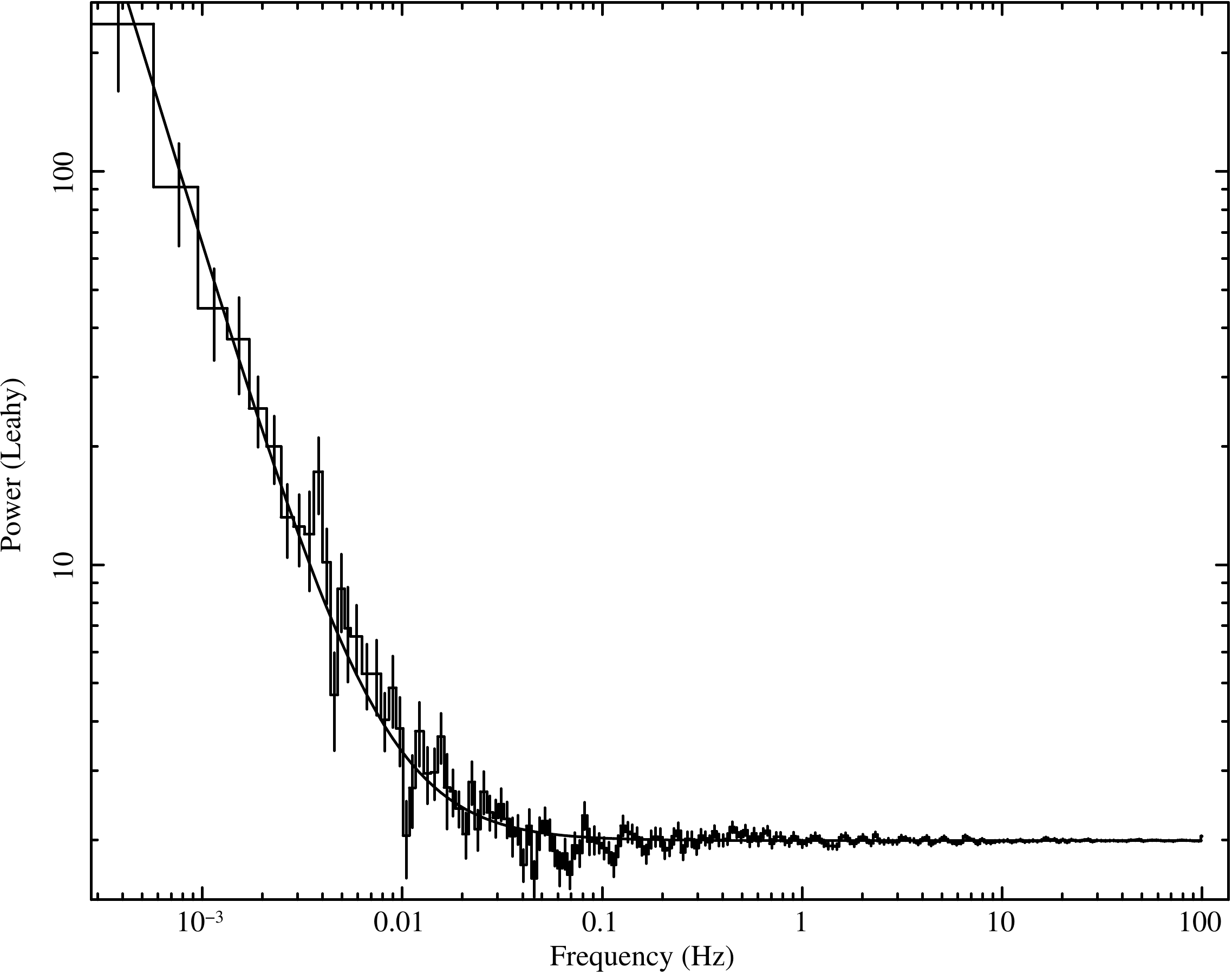}}
	\caption{Power spectral density of the \nustar\ summed light curve in the 3--60\,keV energy range binned at 0.005\,s.
		The power spectrum was obtained by averaging 15 segments of 524288 bins and rebinned geometrically 
		with a factor of 0.05.}
	\label{fig:powspec}
\end{figure}

In order to investigate the origin of the pronounced variability characterizing the \nustar\ lightcurves, we carried out a timing analysis of the 
data by using event files where the arrival time of all recorded photons was framed at the Solar system barycenter using the {\sc barycorr} tool. 
We accumulated light curves in the 3--60\,keV energy range with bins of 0.005\,s and summed FPMA and FPMB to increase the statistics. 
We built a power spectrum averaging on segments of 524\,288 bins
so that most satellite orbits contains one or more intervals. Segments resulting in an exposure of less than 50\%
were discarded, the others padded with zeros. Eventually, we averaged the power spectrum over 15 segments.
In Fig.~\ref{fig:powspec}, we show the resulting power spectrum with the corresponding empirical model function for the combined white and red noise:
a constant ($1.991\pm0.002$) plus a power-law with index $-1.67\pm0.09$. 
 
From the power spectrum of Fig.~\ref{fig:powspec}, we can infer that there is no obvious periodicity. We used the best-fit model of the power spectral density (PSD) as the expected value of the power at each frequency and computed an upper limit on the pulsed fraction between 0.7 and 7\% at 99\% confidence level, using Eq.~(A4) from \citet{israel96}.

Given the relatively low count-rate of the source recorded by \nustar\ and the lack of prominent HR variations in Fig.~\ref{fig:nustar_lc}, we extracted and fit together the FPMA and FPMB spectra of the source accumulated by using the entire exposure time available, after applying an optimal binning \citep{Kaastra2016}.
The source emission is mainly characterized by an absorbed power-law with a high-energy exponential cutoff and an iron line emission centered at $\sim$6.4~keV with equivalent width of about 0.3\,keV. A fit with this simple model left, however, evident residuals especially around 10--20\,keV: an additional component is required to obtain an acceptable result. We have verified that spectral fits with
the most common phenomenological single-component models
leave equivalent residuals (e.g., Fermi-Dirac cutoff, negative and positive power law-NPEX; see \citealt{Coburn2002}). 
We could not  
unequivocally determine whether this required additional component is in absorption or in emission, as both the models \texttt{TBabs*(highecut*pegpwrlaw*gabs + Gaussian)} (referred as \texttt{Gabs} in Table~\ref{tab:spec}) and \texttt{TBabs*(highecut*pegpwrlaw + Gaussian + Gaussian)} (\texttt{Gaussian} in Table~\ref{tab:spec}) could successfully describe the data. Therefore, we report the best-fit parameters obtained by using both models in Table~\ref{tab:spec}. In this table, $N_\mathrm{H}$ is the absorption column density of the \texttt{Tbabs} component, 	$E_\mathrm{C}$ ($E_\mathrm{F}$) is the cut-off (fold) energy of the \texttt{highecut} component, $\Gamma$ is the power-law photon index, and F$_\mathrm{pl}$ is the intrinsic flux of the power-law component. It should be noted that the cutoff energy could not be constrained for the model 
with the additional Gaussian emission feature and was fixed at 1\,keV, outside of the energy range of \nustar\ data.
We indicated with $E_\mathrm{Fe}$, 	$\sigma_\mathrm{Fe}$, and $N_\mathrm{Fe}$ the centroid energy, width, and normalization of the iron line, respectively. $E_\mathrm{Cyc}$, $\sigma_\mathrm{Cyc}$, and $\tau$/$N_\mathrm{Cyc}$ are the centroid energy, width, and depth/normalization of the absorption or emission component, possibly associated with cyclotron scattering.

\begin{table}
	\centering
	\caption{Spectral results obtained by fitting together the \nustar\ FPMA and FPMB data of 
	\igr\ (taken from 2018 August 22 at 11:01 to August 23 at 10:26 UT)). 
	Models and parameters are described in the text.}
	\label{tab:spec}
	\begin{tabular}{l r@{}l r@{}l}
		\hline
		\hline
		& \multicolumn{2}{c}{\texttt{Gabs}} & \multicolumn{2}{c}{\texttt{Gaussian}} \\
		\hline
$N_\mathrm{H}$ [$10^{22}$cm$^{-2}$] & 21 &$^{+2}_{-3}$ & 24 & $\pm$2\\ 
$E_\mathrm{C}$ [keV] & 12.4 & $\pm$0.5 & 1 &--\\ 
$E_\mathrm{F}$ [keV] & 7.8 & $\pm$0.4  & 11 & $^{+2}_{-1}$ \\ 
$\Gamma$ & 0.49 & $\pm$0.12 & 0.51 & $\pm$0.19 \\ 
F$_\mathrm{pl}$ [$10^{-11}$cgs] & 12.0 & $^{+0.4}_{-0.3}$  & 19 & $\pm$3  \\ 
$E_\mathrm{Fe}$ [keV] & 6.32 & $\pm$0.02  & 6.33 & $\pm$0.02 \\ 
$\sigma_\mathrm{Fe}$ [keV] & 0.13 & $\pm$0.07  & 0.12 & $\pm$0.07 \\ 
$N_\mathrm{Fe}$ [$10^{-4}$ph/s/cm$^{-2}$] & 1.5 & $\pm0.2$ & 1.5 & $\pm0.2$ \\ 
$E_\mathrm{Cyc}$ [keV] & 20.1 & $\pm$0.7 & 10.9 & $\pm$0.2 \\ 
$\sigma_\mathrm{Cyc}$ [keV] & 4.1 & $\pm$0.8 & 3.1 & $\pm$0.2 \\ 
$\tau$/$N_\mathrm{Cyc}$ & 4 & $^{+2}_{-1}$ & (1.2 &$\pm0.2)\times10^{-3}$  \\
Flux(0.5--100\,keV)\tablefootmark{a} & \multicolumn{4}{c}{9.6\,\ferg} \\
$\chi^2_\mathrm{red}/d.o.f.$ & 1.026 &/324 & 0.932 &/324 \\ 
\hline
	\end{tabular}
\tablefoot{
	\tablefoottext{a}{Absorbed flux in the 0.5--100\,keV energy range.}
}
\end{table}

\begin{figure*}
	\caption{Unfolded FMPA (black) and FMPB (red) spectra of \igr\ extracted by using the entire exposure time available 
	within the observation ID~90401333002. The plot on the right shows the results for the fit obtained with the model 
	\texttt{TBabs*(highecut*pegpwrlaw + Gaussian + Gaussian)}, while the plot on the left corresponds to the case where the model 
	\texttt{TBabs*(highecut*pegpwrlaw*gabs + Gaussian)} was used. In both cases, the residuals from the fits are shown in the 
	bottom panels.}
	\label{fig:spectra_nustar}
	\includegraphics[width=0.49\textwidth]{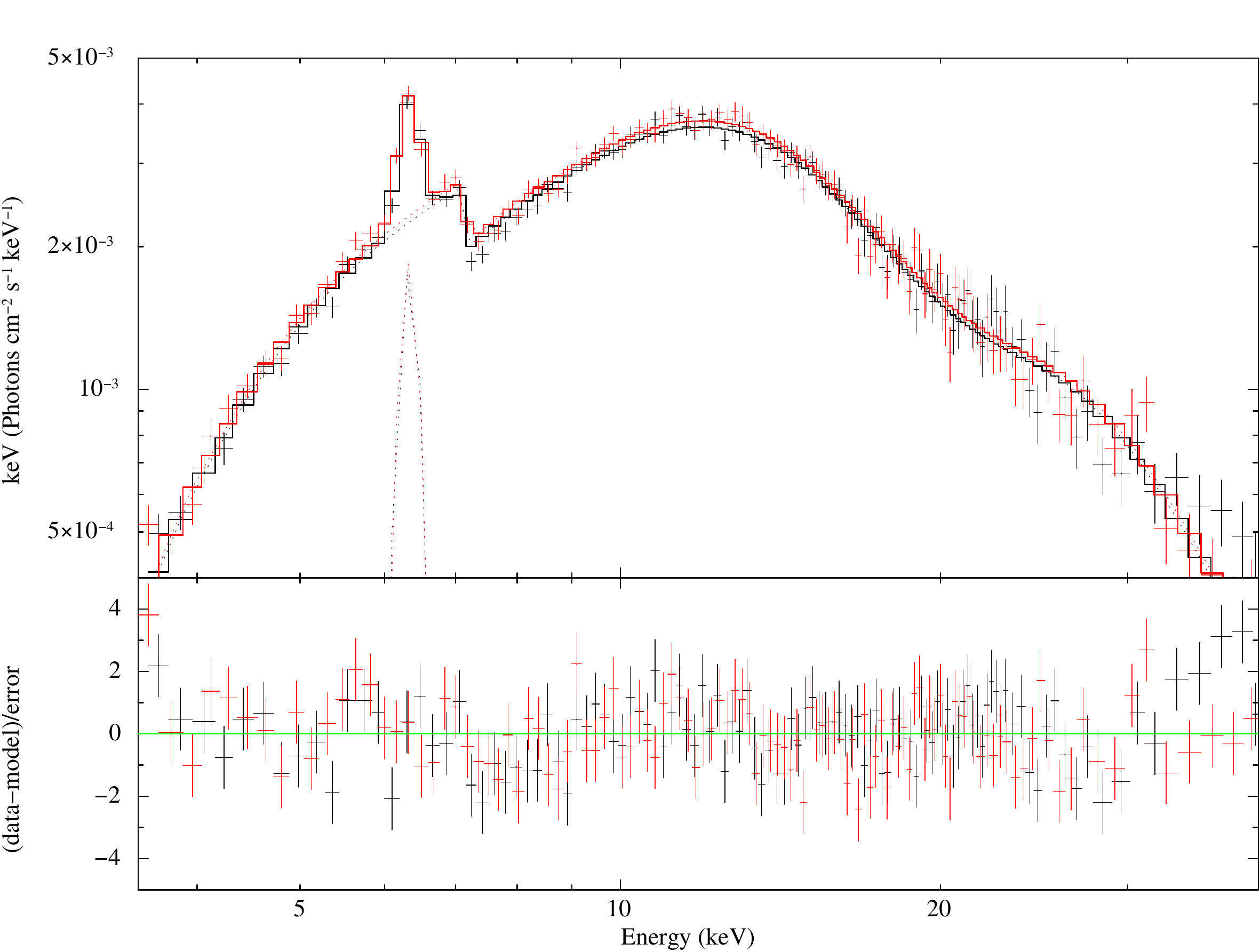}
	\hfill
	\includegraphics[width=0.49\textwidth]{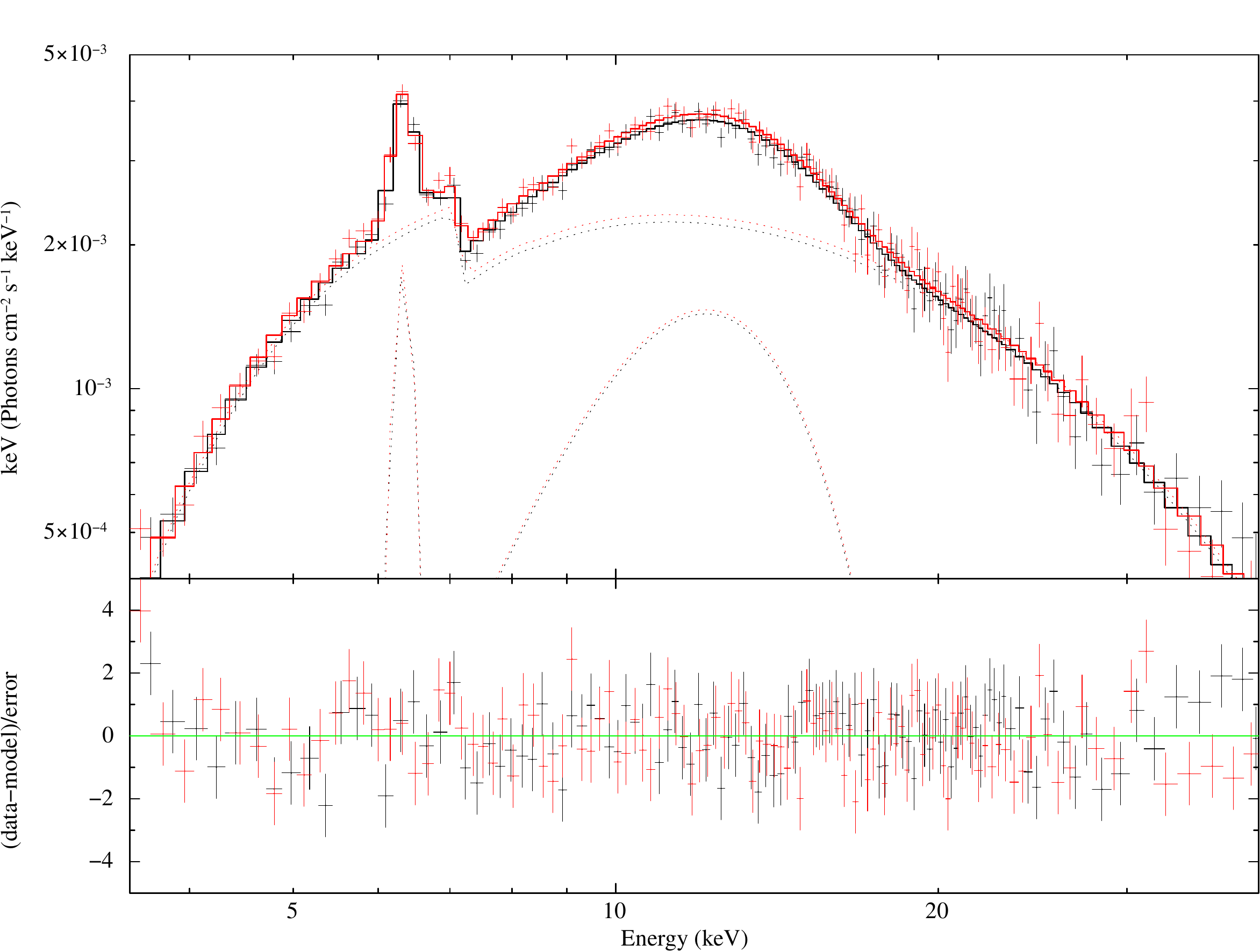}
\end{figure*}

\subsection{\swift\ data}
\label{sec:swift}

\igr\ was observed by the Neil Gehrels \swift\ Observatory \citep{burrows05} twice on 2018 August 13 at 
19:43 (UT) for a total exposure time of 984~s and on 2018 August 23 at 04:39 (UT) for a total exposure time of 1848~s. 
In both cases, data were collected in photon counting (PC) mode.  

\begin{figure}
	\caption{\swift/XRT lightcurve of \igr\ as obtained from the two available observations. The bin time is 100\,s and the lightcurve has been corrected for the 
	background, as well as all instrumental effect (0.5--10\,keV).} 
	\label{fig:xrt_lc}
	
	\resizebox{\hsize}{!}{\includegraphics{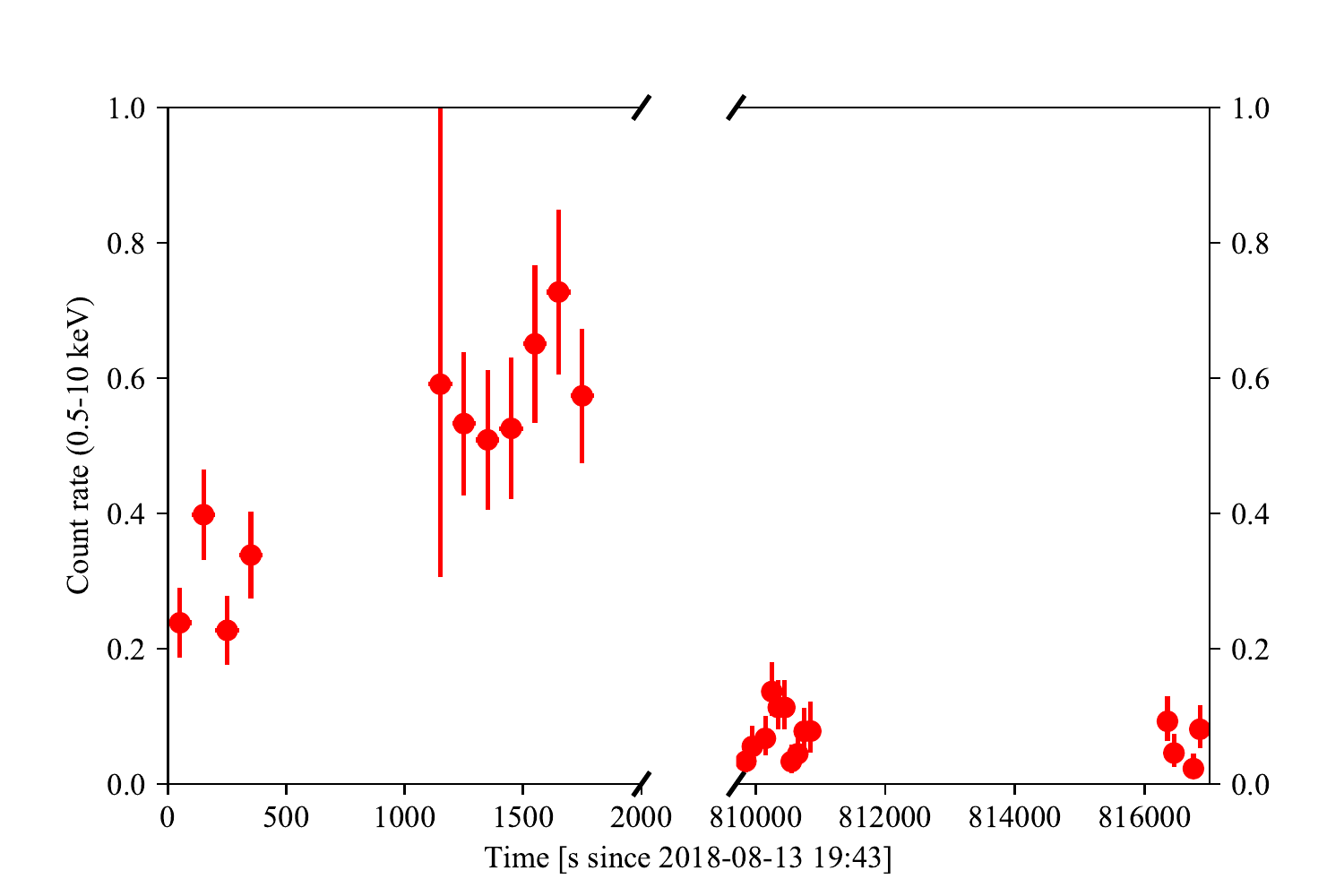}}
\end{figure}

\begin{figure}
	\caption{\swift/XRT spectra extracted from the observations ID 00010807001 (black) and 00088805001 (red). The best fit model is shown with 
		solid lines in the upper panel, while residuals from the best fit are reported in the bottom panel. We used the {\sc xspec} option 
		\texttt{setplot rebin 5 5} for plotting purposes.}
	\label{fig:spectra_swift}
	\resizebox{\hsize}{!}{\includegraphics{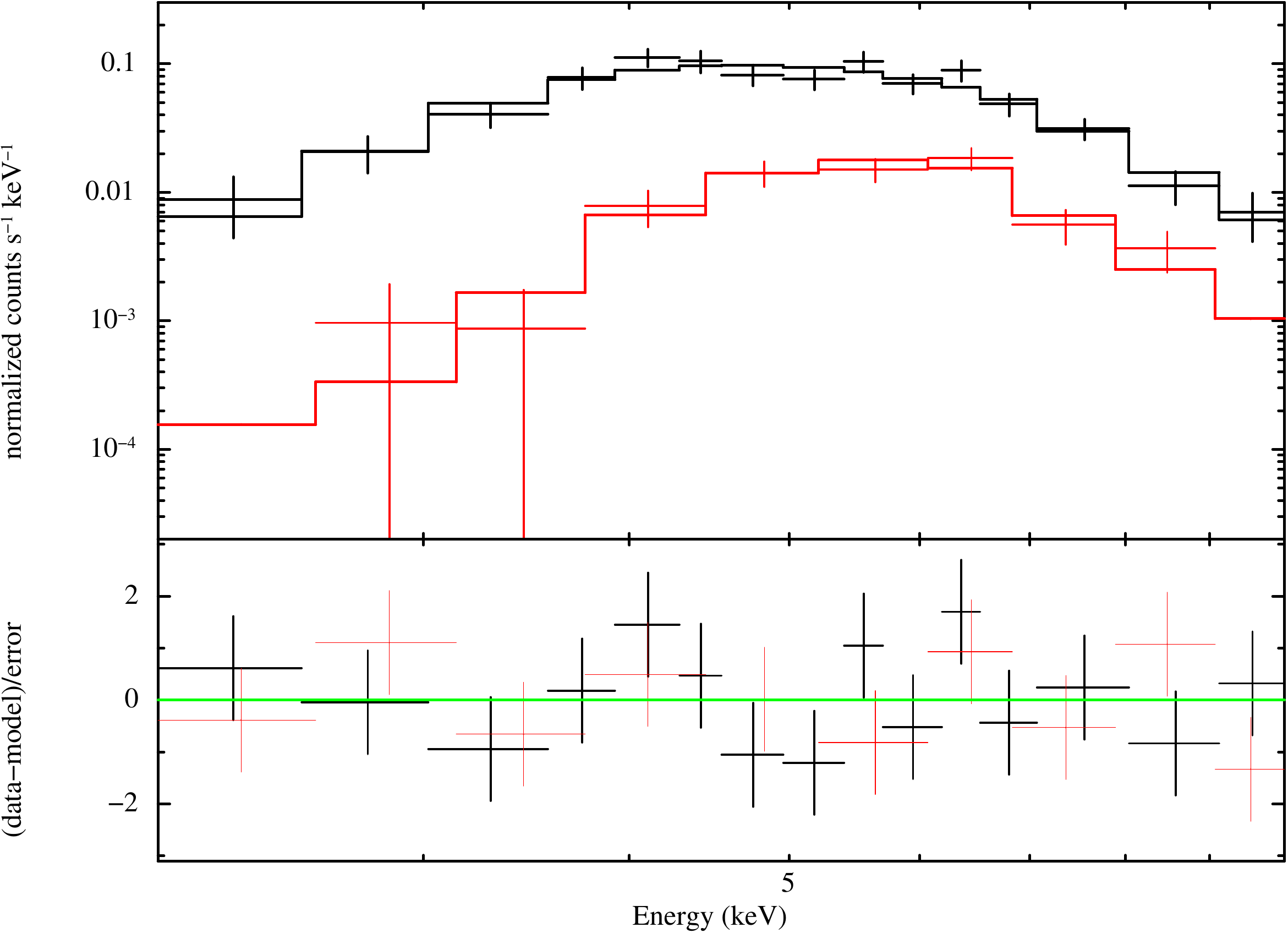}}
\end{figure}

The XRT data were analyzed via the standard software ({\sc Heasoft} v6.25) and the latest  
calibration files available (CALDB 20180710). All data were processed and filtered with {\sc xrtpipeline} (v0.13.4). 
We verified that no data were significantly affected by pileup. The source events were extracted using the photo counting observing mode from a circular region 
with a radius of 20 pixels (where 1 pix corresponds to $\sim2\farcs36$), while 
background events were extracted from a source-free region with a similar radius.   
We show in Fig.~\ref{fig:xrt_lc} the background subtracted XRT lightcurve of the two observations in the 0.5--10~keV energy band 
corrected for point spread function losses and vignetting. The XRT spectra extracted from each observation were binned using the \citet{Kaastra2016} algorithm and
could be well fit with an absorbed power-law model  ({\sc TBabs*pegpwrlaw} in {\sc Xspec}) by minimizing the Cash statistics (C-stat in { \sc Xspec}). A log of XRT observations, together with the corresponding results from the spectral fits, is reported in Table~\ref{tab:summary}. In Fig.~\ref{fig:spectra_swift}, we show the count-rate spectra. It can be noted that during the second observation, the source faded significantly and the spectral parameters are less constrained.

%\begin{table*} 
%\centering 
% \caption{Log and spectral results of the two \swift\/XRT observations used in this paper.}      
% \label{tab:xrt} 
% \begin{tabular}{@{}lcccccccc@{}} 
% \hline 
% \hline 
% \noalign{\smallskip} 
% Sequence   & Obs.  & Start time  & End time & Exposure & $N_{\rm H}$ & $\Gamma$ & Flux$_\mathrm{0.5-10\,keV}$\tablefootmark{a} & $\chi^2_{\rm red}$/d.o.f. \\ 
%                  &   Mode   & (UTC)  & (UTC)  & (s) & (10$^{23}$~cm$^{-2}$) &  & (10$^{-11}$~\ferg) &  (C-stat/d.o.f.)\\
%  \noalign{\smallskip} 
% \hline 
% \noalign{\smallskip} 
%00010807001 & PC & 2018-08-13 19:43 &  2018-08-13 20:13  & 990   &   2.0$^{+1.0}_{-0.6}$  & 0.4$\pm$0.6  &   9.3$\pm$1.5   & 1.1/12 \\
%00088805001 & PC & 2018-08-23 04:39 &  2018-08-23 06:38  & 1848  &   8.0$\pm$5.0  & 0.9$\pm$1.8  &   9.0$\pm$5.0   & (69.7/84) \\
%  \noalign{\smallskip}
%  \hline
%\end{tabular}
%    \tablefoot{
%	\tablefoottext{a}{The indicated flux is in the 0.5--10~keV energy range not corrected for absorption.}}
 
%  \end{table*}

\subsection{\nicer\ data}
\label{sec:nicer}

Following the discovery of \igr,\ Neutron star Interior Composition Explorer \citep[\nicer;][]{nicer} took the opportunity to monitor this source between 2018-08-14 at 21:24 
 and 2018-08-16  13:20 UT with an effective exposure of  $\sim$3.7~ks. We have analyzed data from the observations 
ID 1200320101 and 1200320103 using {\tt NICERDAS} version 2018-04-24 and {\tt HEASoft 6.24} package.  We reprocessed the 
data using the ``nicerl2'' pipeline. GTIs were created using the standard filtering criteria, e.g. the 
angle of bright Earth  $>$40$^{\circ}$, elevation $>$30$^{\circ}$, pointing offset of $\le$54 arcsec and excluding data 
collected during the passages through the South Atlantic Anomaly region. We applied the GTI for the spectral extraction 
using {\tt XSELECT}. \nicer\ observations of blank sky region (\rxte-6) were chosen for background measurements. 
We have used version 1.02 of the \nicer\ response and effective area files for the spectral fitting. 

\begin{figure}
	\caption{\nicer\ spectra extracted from the observations ID 1200320101 (black) and 1200320103 (red). The best fit model is shown with 
	solid lines in the upper panel, while residuals from the best fit are reported in the bottom panel. We used the {\sc xspec} option 
	\texttt{setplot rebin 5 5} and removed empty bins for plotting purposes.}
	\label{fig:nicer}
	\resizebox{\hsize}{!}{\includegraphics{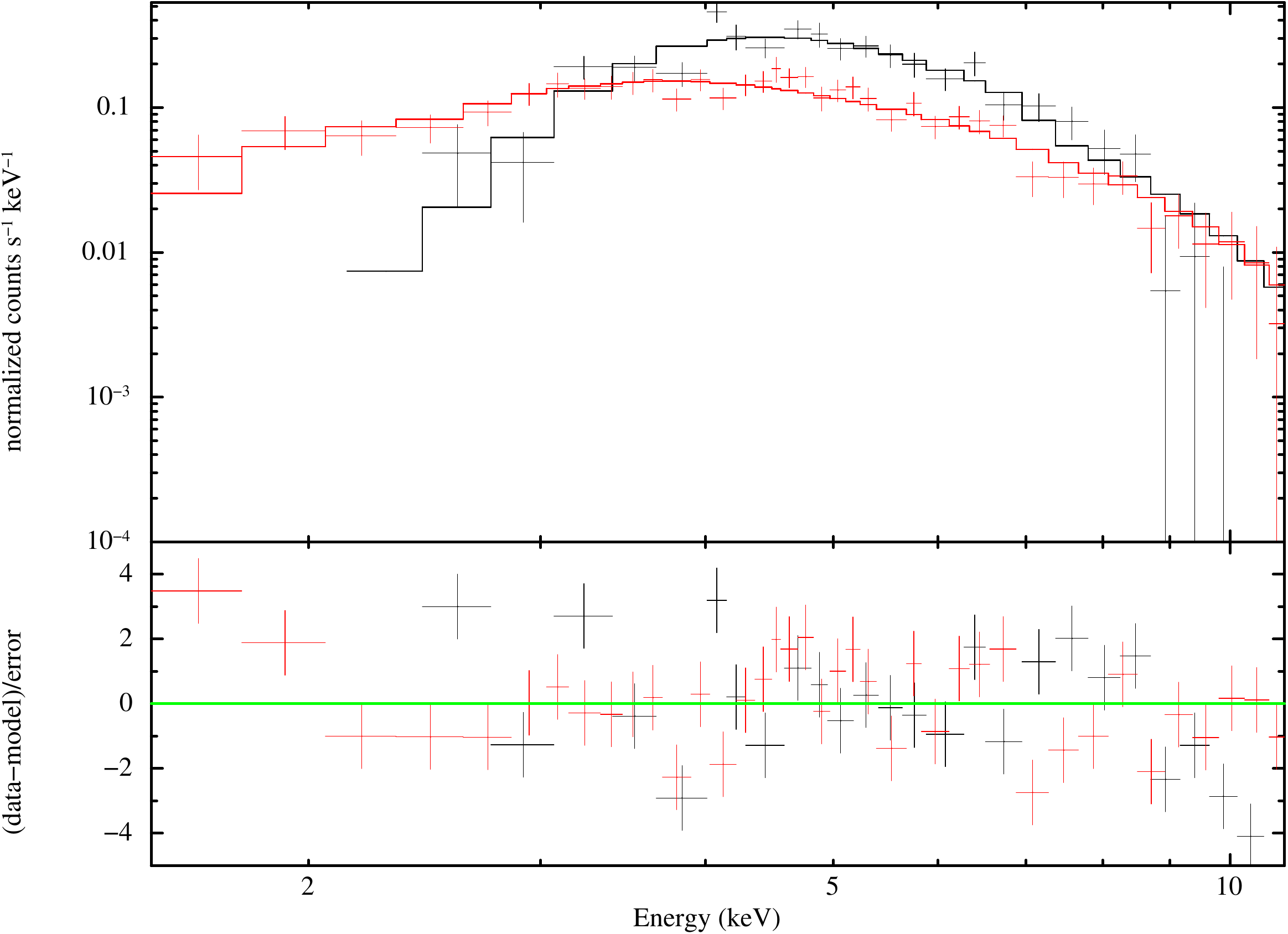}}
\end{figure}

We optimally rebinned the spectra using the algorithm by \citet{Kaastra2016} and limited our energy range from 1 to 11\,keV. Both \nicer\ spectra can be well described with an absorbed power-law model (we used the model {\sc TBabs*pegpowerlaw}). We report the results in 
 Table~\ref{tab:summary} noting that we minimized the Cash statistic with background subtraction (C-stat in {\sc Xspec}). The two spectra, together with the residuals from the best fit, are shown in Fig.~\ref{fig:nicer}.
 
 The power density spectrum of the \nicer\ observation is consistent with white noise from $\sim10^{-3}$ to 500\,Hz. The
 	upper limits we could obtain on the pulsed fraction are less containing than those reported for
 	\nustar\ data owing to the shorter exposure and high background of the \nicer\ observations. 
%\begin{table*} 
%\centering 
% \caption{Results obtained from \nicer\ data. 
% 	The model is an absorbed power law model ({\sc TBabs*pegpwrlaw} in {\sc Xspec)}.}      
% \label{tab:nicer} 
% \begin{tabular}{@{}lccccccc@{}} 
% \hline 
% \hline 
% \noalign{\smallskip} 
%ID & TSTART & TSTOP& Exposure & $N_{\rm H}$ & $\Gamma$ & Flux$_{\rm 0.5-10\,keV}$\tablefootmark{a} &C-stat/d.o.f. \\ 
% & (UTC) & (UTC) & (s) & (10$^{22}$~cm$^{-2}$) &  & (10$^{-11}$~\ferg) & \\
%  \noalign{\smallskip} 
% \hline 
% \noalign{\smallskip} 
%1200320101 & 2018-08-14 21:24 & 2018-08-14 22:43 & 761 & 26$^{+6}_{-5}$  & 2.4$^{+0.6}_{-0.7}$  &   22$^{+50}_{-12}$ & 175/131\\
%1200320103 & 2018-08-16 02:14 & 2018-08-16 13:20 & 2888 & 5.0$^{+1.5}_{-1.2}$  & 0.4$\pm$0.3  &   1.7$^{+0.17}_{-0.14}$ & 114/138\\
%  \noalign{\smallskip}
%  \hline
%  \end{tabular} 
%    \tablefoot{
%	\tablefoottext{a}{The indicated flux is in the 0.5--10~keV energy range, corrected for absorption.}}
%  \end{table*}

\section{Discussion}
\label{sec:discussion}

We reported on the transient source \igr,\ whose first detectable X-ray emission was discovered with \inte\ on 2018 August 11. The source was followed up by \chan,\ \nustar,\ \swift,\ and \nicer.\ 

\igr\ was relatively faint for the instruments on-board \inte,\ and thus only a preliminary description of the broad-band spectrum could be obtained from the available JEM-X and IBIS/ISGRI data. \inte\ data showed that the X-ray emission from the source faded very rapidly, decreasing below the detection limit of JEM-X and IBIS/ISGRI within one revolution (2.7 days), supporting a fast flaring behavior. The more sensitive instruments on-board \nustar,\ \swift,\ and \nicer\ revealed that significant X-ray emission from the source 
could be detected at least until early August 23, albeit at an intensity about a factor of $\sim$50 lower than that recorded by \inte.\ The \chan\ observation carried 
out later during the same day showed a further decrease in intensity by at least another factor of $\sim$10 and down to 2.9$\times$10$^{-12}$~\ferg 
\citep{chakrabarty18, chakrabarty18b}. 

In the soft to hard X-ray domains (0.5--80\,keV), the emission from \igr\ is characterized by a remarkable variability on time scales from a few seconds to a few 
thousands of seconds, typical of what is usually observed in wind-fed high mass X-ray binaries \citep[HMXBs; see][for a recent review]{walter15}. This variability is 
best appreciated by looking at the \nustar\ lightcurves which are endowed with a much higher S/N and longer exposure, compared to all other available instruments 
(see Fig.~\ref{fig:nustar_lc}). Although within the \nustar\ observation we did not record prominent variations in the HR, a comparison between all spectra results\footnote{Note that we did not attempt to perform a combined fit between the different instruments because data from \inte,\ \nustar,\ \nicer,\ and \swift\ were obtained at largely different epochs and the significant spectral changes recorded in these data would make it very difficult to provide a consistent  interpretation of the results. Furthermore, given the large absorption column density revealed in the direction of the source, the limited S/N of the \nicer\ and \swift\ data would not add significant information to the fit of the \nustar\ data that can satisfactorily cover alone a broad-band emission range extending from 3 up to 40~keV.} 
reported in Sect.~\ref{sec:integral}-\ref{sec:nicer} highlights significant changes in the absorption column density (up to a factor of $\sim$10) and in the 
power-law slope (up to a factor of $\sim$3). These changes, as well as the much higher value of $N_{\rm H}$ compared to the 
expected Galactic extinction in the direction of the source \citep{chenevez18}, supports the idea of \igr\ being a wind-fed HMXB. It is well known that 
in these systems the fast wind of a massive companion star (typically an OB supergiant) can lead to the formation of a cocoon of dense material around 
the accreting compact object and the local density/velocity variations of the wind can give rise to an X-ray variability on compatible time scales compared to those 
observed from \igr\ \citep[see, e.g.,][for a recent review]{nunez17}. 

Although the present observations do not allow to firmly establish the nature of the accreting compact object in this source, the broad-band spectral analysis revealed properties that are strongly reminiscent of what is usually observed from neutron star HMXBs \citep[see, e.g.,][]{klochkov2007, walter2015}. The cut-off power-law spectrum is commonly observed in these systems, and the iron line at 6.4~keV is often observed as a consequence of the fluorescence of X-rays from the neutron star (NS) onto the surrounding stellar wind material or accretion disk. The peculiar feature around 10--20\,keV that we modeled with either a Gaussian emission component or a multiplicative absorbing Gaussian profile is a known signature of a strongly magnetized NS. The interpretation of an emission feature at these energies is associated, as suggested for other HMXBs, to thermal and bulk Comptonization of magnetized bremsstrahlung seed photons along the accretion column \citep{ferrigno09, farinelli16}. The presence of an absorption feature would be explained by assuming this is a broad absorption line produced by scattering 
on electrons bounded around the field lines of a strong magnetic field (also known as ``cyclotron line'). These features are observed in many NS HMXBs and their centroid energy provides a direct estimate of the NS magnetic field strength according to the equation: 
\begin{equation}
E_{cyc} = \frac{1}{(1+z)}  \frac{\hbar eB}{m_e c} \approx \frac{1}{(1+z)} 11.6  \times B_{12}\,\mathrm{keV},
\end{equation}
where $B_\mathrm{12}$ is the magnetic field in units of $10^{12}$\,Gauss and $z$ is the gravitational 
redshift at the emission site \citep[the above equation is valid for the fundamental CRSF; see, e.g.][for a recent review]{staubert18}. If we accept this interpretation, we can conclude that the accreting object in \igr\ is a NS endowed with a magnetic field of $\sim$2$\times$10$^{12}$~G, compatible to other measured NS magnetic field strengths in HMXBs. The 4\,keV width of the scattering feature is also typical for these objects
and might indicate a plasma temperature of 10--20\,keV, in agreement with the cutoff energy. However, this
should be taken with caution, since the line width depends on the line-of-sight angle to the magnetic field.
Future observations of additional episodes of enhanced X-ray emission from \igr\ with a high sensitivity broad-band X-ray instrument as the FPMs on-board NuSTAR will hopefully be able to provide higher S/N spectra and rule out any other alternative interpretation of the feature around 20~keV from the spectral fitting.  

A final convincing indication of the HMXB nature of \igr\ is provided by the identification of the near-infrared counterpart reported by \citet{masetti18}. 
This led to the specific association of the source with the class of the supergiant HMXBs\footnote{Note that other authors have questioned this conclusion, even though a firm identification of this object might require additional observations \citep{collum18}.}. 
These systems are generally divided into two sub-classes, the so-called classical systems and the supergiant fast X-ray transients \citep[SFXTs; see, e.g.,][for a recent review]{nunez17}. 
The former are variable but persistent systems, showing on average a luminosity that is well explained by using a wind accretion scenario onto a compact object (usually a strongly magnetized neutron stars). 
The SFXTs display a much prominent variability in the X-rays, alternating between hours-long outbursts reaching the typical luminosity of classical systems and extended periods of quiescence where the X-ray luminosity can decrease by up to a factor of 10$^5$--10$^6$. 
This behavior is far from being understood and it is still actively debated \citep[see, e.g., the discussion in][]{bozzo17}. 
Data collected so far from \igr\ would favor the connection with the SFXT sub-class, as the source underwent a relatively bright episode of X-ray emission at the time of the discovery with JEM-X and then progressively fainted down until it got close to the detection threshold also for \chan\ \citep[the X-ray bright phase displayed by \igr\ was about $\sim$12~days long in total, a duration that is not uncommon in other SFXTs; see, e.g.,][]{sguera15}. 
Although the total variation in the X-ray luminosity recorded so far is of $\sim$300 (see Sect.~\ref{sec:intro}) and thus significantly lower than that usually measured from the SFXTs, the fact that no X-ray emission was ever recorded before from this object suggests that the true quiescent luminosity could be lower than the value measured during the \chan\ observation. The overall dynamic range in the X-ray domain could, therefore, be even larger. 
\citet{masetti18} suggested from the reddening of the OB supergiant that this system is located beyond the Galactic Center at $\sim$10~kpc, and thus the outburst luminosity derived from the JEM-X flux would be of
$\sim$2$\times$10$^{36}$\,erg\,s$^{-1}$, while the \nustar\ 0.5--100\,keV flux would correspond to a luminosity of $\sim$10$^{36}$\,erg\,s$^{-1}$.
This value is similar to what is observed from the faintest SFXT outbursts, that can achieve a luminosity up to $\gtrsim$10$^{38}$~\ferg \citep{romano15b}. 
As a consequence, it is possible that this is a peculiarly faint system also during outbursts, explaining why no previous detection with \inte\ or \swift/BAT was ever reported whilst usually up to few outbursts per year are detected from the known SFXTs \citep[see, e.g.,][]{paizis14,romano15,sidoli18}.

\section*{Acknowledgements}
We are grateful to the \nustar,\ \swift,\ and \nicer\ teams for the prompt scheduling of the ToO observations on 
\igr.\
We acknowledge financial contribution from the agreement ASI-INAF n. 2017-14-H.O.
We acknowledges support from the HERMES Project, financed by the Italian Space Agency (ASI) Agreement n. 2016/13 U.O. GKJ acknowledges support from 
the Marie Sk{\l}odowska-Curie Actions grant no. 713683 (H2020; COFUNDPostdocDTU)
 %We thank the anonymous referee for detailed comments that helped us improve the paper. 

\bibliography{reference.bib}{}
\bibliographystyle{aa}

\end{document}